\documentclass[twocolumn,pre,nobibnotes]{revtex4}
\usepackage{graphicx}
\usepackage{epstopdf,graphics,subfigure,psfrag,amsmath,amssymb}

\begin{document}

\title{Chimera dynamics in nonlocally coupled moving phase oscillators}
\author {Wenhao Wang, Qionglin Dai, Hongyan Cheng}
\author{Haihong Li}\email{haihongli@bupt.edu.cn}
\author{Junzhong Yang}\email{jzyang@bupt.edu.cn}
\address{$^1$School of Science, Beijing University of Posts and
Telecommunications, Beijing, 100876, People's Republic of China}
\date{\today}

\begin{abstract}{Chimera states, a symmetry-breaking spatiotemporal pattern in nonlocally coupled dynamical units, prevail in a variety of systems. However, the interaction structures among oscillators are static in most of studies on chimera state. In this work, we consider a population of agents. Each agent carries a phase oscillator. We assume that agents perform Brownian motions on a ring and interact with each other with a kernel function dependent on the distance between them. When agents are motionless, the model allows for several dynamical states including two different chimera states (the type-I and the type-II chimeras). The movement of agents changes the relative positions among them and produces perpetual noise to impact on the model dynamics. We find that the response of the coupled phase oscillators to the movement of agents depends on both the phase lag $\alpha$, determining the stabilities of chimera states, and the agent mobility $D$. For low mobility, the synchronous state transits to the type-I chimera state for $\alpha$ close to $\pi/2$ and attracts other initial states otherwise. For intermediate mobility, the coupled oscillators randomly jump among different dynamical states and the jump dynamics depends on $\alpha$. We investigate the statistical properties in these different dynamical regimes and present the scaling laws between the transient time and the mobility for low mobility and relations between the mean lifetimes of different dynamical states and the mobility for intermediate mobility.}
\end{abstract}


\maketitle

\section{Introduction}
Chimera states refer to the type of spatiotemporal pattern consisting of coexisting coherent and incoherent regions. Since the first observation in nonlocally coupled identical phase oscillators in 2002 \cite{kura}, chimera states have evolved in the last decade from surprising symmetry-breaking patterns to prevailing dynamical phenomena ranging from physics and chemistry to biology, from classical to quantum systems \cite{abra,lai09,mot10,zhu12,pan15,mart13,tins12,hage12,cheng18,gav18,lai18}. Chimera states are not only numerically found in mathematical models such as periodic and chaotic maps \cite{omel11}, mechanical oscillators \cite{omel15}, neuronal oscillators \cite{omel13,hiz14,sak06}, but also experimentally realized in mechanical systems \cite{mart13,oml15}, optical systems \cite{hage12} and chemical systems \cite{tins12}. Recent works have shown that chimera states may exist even in oscillators coupled globally \cite{yel14,set14,cha14,pre15} as well as locally \cite{lai15,ber16}. Chimera states are not restricted to topologies such as rings, square lattices \cite{mart10a,gu13}, torus \cite{pana13}, spheres \cite{pana15}, and Erd\"{o}s-R\'{e}nyi networks and scale-free networks \cite{zhuy14}. Chimera states are robust to random rewiring of edges \cite{jiang16} in a ring of symmetrically nonlocally coupled phase oscillators. Though most of works dealt with coupled oscillators where units support self-oscillating solutions, chimera states have been studied in nonlocally coupled
excitable systems only allowing for an equilibrium. Semenova et. al. realized chimera state in excitable units in the presence of noise through coherent resonance \cite{sem16} and Dai et. al. found that chimera state may emerge out of excitable units through a coupling-induced collective oscillation \cite{dai18}. The connections between chimera states and other dynamical phenomena have been investigated. Motter et. al. proposed a connection between chimera states and cluster synchronization in networks of locally coupled chaotic oscillators \cite{mot17}. Lai et. al. established a connection between chimera states and a quantum scattering phenomenon in 2-dimensional Dirac material systems where manifestations of classically integrable and chaotic dynamics coexist simultaneously \cite{lai18}. Dai et. al. reported a desynchronization transition from synchronous to asynchronous chimera states in a bicomponent phase oscillator system when the frequency mismatch among oscillators increases \cite{dai18b}.

Most of previous works on chimera states have assumed the interaction between oscillators to be static. However, it has been found in various fields, such as power transmission system \cite{sac00}, consensus problem \cite{olf07}, and person-to-person communication \cite{onn07}, that there are many interesting scenarios while interaction patterns are time-varying. The time-varying interaction patterns may be realized either by rewiring connections in networks in the course of time or by assuming oscillators to move in space. In the past few years, the interest in the collective dynamics in coupled oscillators has grown rapidly when oscillators may move in space. Different dynamical behaviors have been investigated, such as amplitude death and resurgence of oscillation \cite{maj17}, mobility-enhanced signal response \cite{shen13}, and synchronization \cite{fra08,gom13,eom16}. It has been shown that synchronization time may depend non-monotonically on the mobility of oscillators and the mechanism of driving synchronization are different for different dynamical regimes \cite{fuj11,pri13,bea17}. Chimera dynamics in mobile oscillators has not been investigated except for a recent work, in which an array of locally coupled oscillators exchange positions and the persistent chimera states may be maintained under the interplay between the mobility and time delay \cite{Petr17}.

In this paper, we study how chimera dynamics reacts to the mobility in a ring of nonlocally coupled phase oscillators and identify different regimes ruling the chimera dynamics. The rest of paper is arranged as follows. In section 2, we present the model in which $N$ agents, each of which is associated with a phase oscillator, perform Brownian motions on a ring. In section 3, we first present the dynamical states existing in the model when agents are motionless. Then we study the response of these dynamical states to the mobility and the statistical properties of the responses. Finally, we conclude with a summary in section 4.\emph{}

\section{Model}

We consider $N$ agents performing independent Brownian motions on a ring with the length $L$. The position of agent $i$, $x_i$, evolves according to
\begin{eqnarray}\label{eq1}
\dot{x}_{i}(t)&=\eta_i(t)
\end{eqnarray}
with periodic boundary condition $x_i(t)+L=x_i(t)$. We assume that the noise $\eta_i(t)$ has a Gaussian probability distribution with mean and correlation function
\begin{eqnarray}\label{eq2}
\langle\eta_i(t)\rangle&=&0,\nonumber\\
\langle\eta_i(t)\eta_j(t')\rangle&=&D\delta_{ij}\delta(t-t').
\end{eqnarray}
$D$ denotes the noise strength.

Furthermore, each agent is associated with a phase oscillator and, then, each agent is characterized by a phase variable $\theta$. Agents interact with each other with a nonlocal coupling strength depending on the distance between them and the phase variable $\theta_i$ of agent $i$ evolves in terms of the following equation
\begin{eqnarray}\label{eq3}
\dot{\theta}_{i}(t)&=&\omega-\sum_{\text{j=1}}^{\text{N}}G(d_{ij}(t))\sin[\theta_{i}(t)-\theta_{j}(t)+\alpha].
\end{eqnarray}
$\alpha$ is the phase lag and $\omega$ is the natural frequency for all oscillators. Without losing generality, $\omega$ is set to be zero. $d_{ij}(t)=min(|x_i(t)-x_j(t)|,L-|x_i(t)-x_j(t)|)$ is the distance between agents $i$ and $j$. The kernel function $G(d_{ij}(t))$, the nonlocal coupling between oscillators $i$ and $j$, is assumed to be even, nonnegative, decreasing with $d_{ij}(t)$. To be concrete, we consider widely used kernel function $G(d_{ij}(t))=[1+A\cos(2\pi d_{ij}(t)/L)]$ with $0\leq A\leq1$. Comparing with the exponential kernel $G(x)\propto exp(\kappa|x|)$ \cite{kura}, the cosine kernel allows the model to be solved analytically besides qualitatively similar results\cite{abra}. The parameter $A$ controls the nonlocal coupling, and set to be $1$ in this paper. The other parameters are chosen as $N=128$ and $L=2\pi$. To our knowledge, the model (3) with all  motionless oscillators evenly distributed along the ring allows for two types of chimera states besides the synchronous state. There is only one coherent domain in the type-I chimera while there are two coherent domains in the type-II chimera with oscillators in different coherent domains being in antiphase. For $A=1$, the type-I chimera is stable for $\alpha$ higher than around $1.31$ \cite{abra} while the type-II chimera exists for $\alpha$ higher than around $1.43$.

\begin{figure}
\includegraphics[width=3.5in]{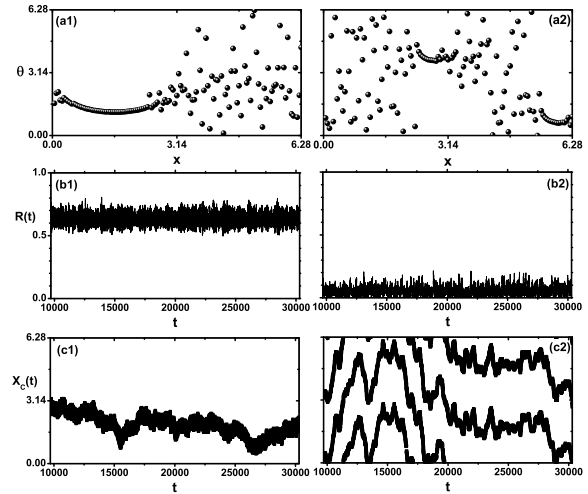}
\caption{\label{fig1}The chimera states at $\alpha=1.38$ in the left column and at $\alpha=1.47$ in the right column. (a) The snapshot of $\theta$. (b) The evolution of $R(t)$. (c) The evolution of the locations of the coherent clusters $X_C(t)$. $D=0$ and oscillators are assigned with positions $x_i=iL/N$ ($i=1,2,\cdots,N$).}
\end{figure}

\section{Results and Discussion}

We consider three types of initial conditions prepared at the corresponding $\alpha$ with the positions of motionless agents $x_i=iL/N$ ($i=1,2,\cdots,N$), the synchronous state with all oscillators in phase [$\theta_i(t)=\theta(t)$], the type-I chimera with one coherent domain in which coherent oscillators are almost in phase(shown in left column of Fig.~\ref{fig1}), and the type-II chimera with two coherent domains (shown in right column of Fig.~\ref{fig1}). The synchronous state is always locally stable when $\alpha<\pi/2$ while chimera states require $\alpha$ close to $\pi/2$ to be stable. In Fig.~\ref{fig1}, we present the type-I chimera at $\alpha=1.38$ and the type-II chimera at $\alpha=1.47$. The difference of the two chimera states in the number of coherent domains is obvious in the snapshots of the spatiotemporal plots of oscillators' phases. To further distinguish the two chimera states, we consider two quantities. The first is the global order parameter defined as $Z(t)=R(t)e^{i\Theta(t)}=\sum_{i=1}^{N}e^{i\theta_i(t)}$. $R(t)$, the amplitude of global order parameter, fluctuates around $0.65$ for the type-I chimera. In contrast, $R(t)$ for the type-II chimera is much lower since oscillators in different coherent domains are in antiphase, as shown in Fig.{1(b2)}, $R(t)$ is always less than $0.35$ for the type-II chimera. The second quantity is the location of the coherent domains, $X_c$, represented by the center of the coherent domain. To measure it, we consider position-dependent order parameter $Z(x,t)=R(x,t)e^{i\Theta(x,t)}=\sum_{\text{j=1}}^{\text{N}}G(d_j)e^{i\theta_{j}(t)}$ with $d_{j}=min(|x-x_j|,L-|x-x_j|)$. $R(x,t)$ roughly reaches its local maximum at $X_c$. The locations of the coherent domains, $X_c$, are defined as the positions at which the amplitude $R(x)$ reaches its local maximum. Since there are two coherent domains in the type-II chimera, as shown in Fig.~\ref{fig1}(c2), the coherent domains for these two chimera states move along the ring in an irregular way, which is due to the finite size effects \cite{wol11}. It has been concluded theoretically for the type-I chimera that the size of the coherent domain decreases with $\alpha$ increase \cite{abra}, which suggests that the global order parameter $R$ decreases with $\alpha$. It can be numerically confirmed that the type-II chimera against $\alpha$ behaves in a similar way.

Then we consider how the model Eqs.~(\ref{eq1}-\ref{eq3}) reacts to the movement of agents. With nonzero $D$, agents walk randomly along the ring. Between successive times $t$ and $t'$, the displacement of an agent , $|x(t)-x(t')|=\sqrt{2D(t-t')}$ is given as square root for short time interval $t-t'$, which suggests that the mobility of agents may be measured by the noise strength $D$ and, large $D$ means high mobility. The relative positions between agents change as time goes on, which takes effects on the dynamics of agents in two aspects. Firstly, the mobility of agents may act as perpetual perturbation to chimera states when coherent oscillators may move out of coherent domains and incoherent oscillators may move into coherent domains. The higher the mobility, the stronger the perturbation. It is well known that, for the model here, chimera states in a finite number of motionless agents are always transient to the synchronous state \cite{wol11} but with transient time increasing with the number of oscillators in an exponential way. Therefore, the mobility of agents may drive a chimera state to the synchronous state. Secondly, the summation $\sum_{\text{j=1}}^{\text{N}}G(d_{ij})$ is no longer a constant and, it becomes both time-dependent and agent-dependent. Consequently, the time-varying summation for each agents actually act as a perpetual perturbation to the synchronous state, may prevent chimera states from being attracted by the synchronous state.

\begin{figure}
\includegraphics[width=3.5in]{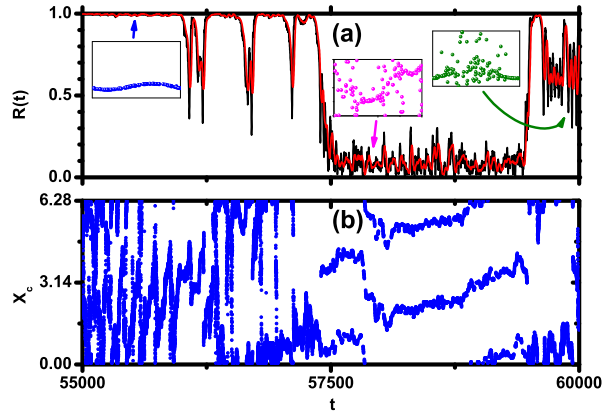}
\caption{\label{fig2}(a) The evolutions of $R(t)$ (black) and the window averaged $R(t)$ (red). The insets show the snapshots of $\theta$ for different dynamical states, the synchronous state (blue) where $R$ close to 1, the type-I chimera (green) with $R$ close to $0.65$, and the type-II chimera (pink) with $R$ close to $0$. The arrows denote the time at which snapshots are taken. (b) The evolution of the locations of the coherent domains. $A=1$, $\alpha=1.38$, and $D=4\times10^{-3}$.}
\end{figure}

For the convenience of illustration in the following, we present an example of dynamical process with moving agents. Fig.~\ref{fig2}(a) shows that $R$ spends most time on the sections with $R$ close to 1, $R$ close to 0, and $R$ close to $0.65$ and these sections are connected by frequent jumps among them. The insets in Fig.~\ref{fig2}(a) show the snapshots of oscillators in these three time sections with different $R$. Clearly, the sections with $R$ close to $1$ is in the synchronous state while the time sections with $R$ close to $0.65$ and $0$ are featured by the type-I and the type-II chimeras, respectively. We also plotted the averaged $R$ in the moving window with the length of $50$ time units in order to reduce the fluctuation in $R$ in value and the qualitative features are the same. Figure~\ref{fig2}(b) plots $X_c$ with time. In the time sections of the type-II chimera, there are two possible $X_c$ at the same time. In contrast, the synchronous state and the type-I chimera state are hard to be distinguished by $X_c$ though $X_c$ varies slowly with time in the sections.

\begin{figure*}
\includegraphics[width=6in]{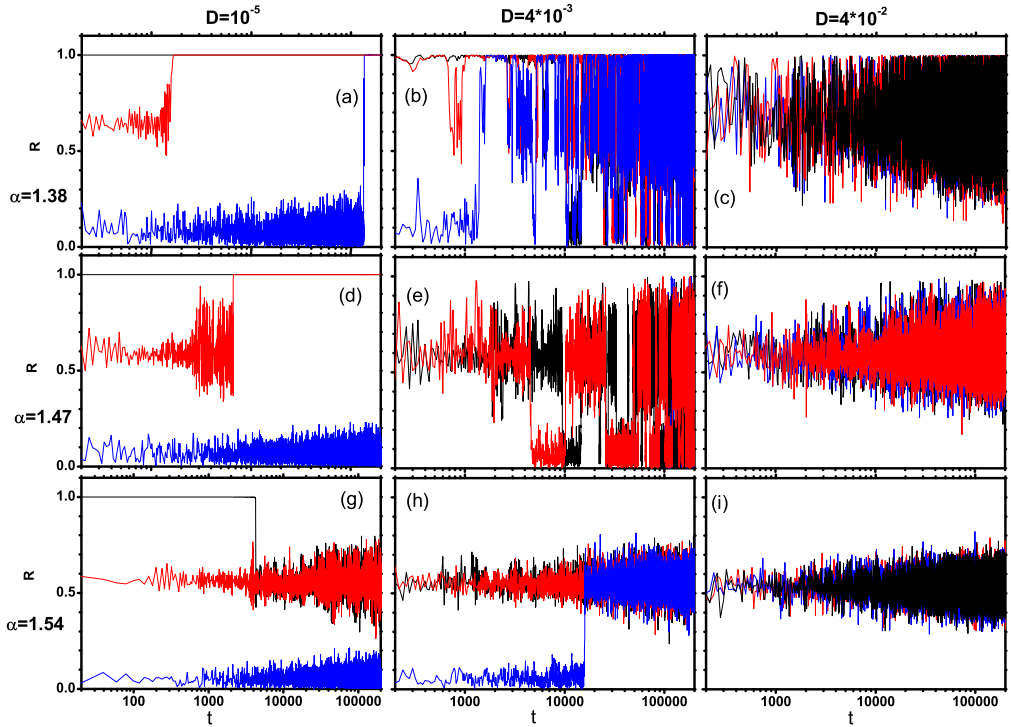}
\caption{\label{fig3}The evolutions of $R$ starting with the synchronous state (black), the type-I chimera (red), and the type-II chimera (blue) for different $\alpha$ and $D$. $D=10^{-5}$ in the left column, $D=4\times10^{-3}$ in the middle column, and $D=4\times10^{-2}$ in the right column. $\alpha=1.38$ in the top row, $\alpha=1.47$ in the middle row, and $\alpha=1.54$ in the bottom row.}
\end{figure*}

The response of the model Eqs.~(\ref{eq1}-\ref{eq3}) to the mobility of agents depends on $\alpha$. We consider three cases: 1) $\alpha=1.38$ close to the stability boundary of the type-I chimera where the type-II chimera cannot be maintained; 2) $\alpha=1.47$ where both the type-I chimera and the type-II chimera can be maintained; 3) $\alpha=1.54$ close to $\alpha=\pi/2$ beyond which incoherent state is always stable. We consider three types of initial conditions, the synchronous state and the type-I chimera prepared at the corresponding $\alpha$ with the positions of motionless agents $x_i=iL/N$, and the type-II chimera in Fig.~\ref{fig1}(a2). For each parameter combination, $\alpha$ and $D$, and each type of initial conditions, we run tens of  realizations and the typical results are presented in Fig.~\ref{fig3}. The left column in Fig.~\ref{fig3} shows the results for weak mobility $D=10^{-5}$. For  $\alpha=1.38$, both the type-I and the type-II chimeras evolve towards the synchronous state though the transient time for the type-II chimera is much longer than that for the type-I chimera. However for $\alpha=1.47$, only the type-I chimera will transit to the synchronous state while the type-II chimera is maintained for the simulation time up to $6\times10^{5}$ time units. Interestingly, weak mobility may transfer the synchronous state to the type-I chimera at $\alpha=1.54$ while the type-II chimera is robust to the perturbation induced by the mobility of agents. For intermediate mobility $D=4\times10^{-3}$, the coupled oscillators response to the mobility of agents at $\alpha=1.38$ by jumping among the synchronous states, the type-I and the type-II chimeras in an irregular way regardless of the initial states. In contrast, the coupled oscillators jump only between the type-I and the type-II chimeras at $\alpha=1.47$. The random jumps between the two chimeras are lost at $\alpha=1.54$. The right column in Fig.~\ref{fig3} shows the situation with high mobility $D=4\times10^{-2}$. Independent of $\alpha$ and initial conditions, $R$ fluctuates around the value close to the type-I chimera. Increasing $\alpha$ reduces the fluctuation. Actually, there is no clear signature for the type-I chimera to be identified, which may be caused by fast movement of $X_c$ if the dynamical state does belong to the type-I chimera.

\begin{figure}
  \centering
  \includegraphics[width=3.5in]{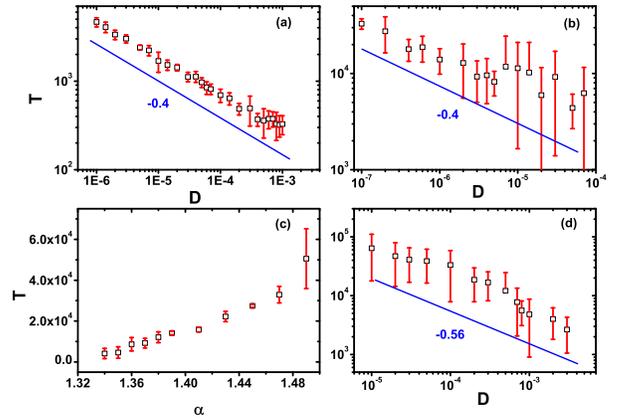}
  \caption{\label{fig4} The transient time $T$ is plotted against the mobility $D$ at $\alpha=1.38$ in (a) and at $\alpha=1.47$ in (b), respectively, with the type-I chimera as the initial state. The blue line denotes the fitting curve $T\sim D^{-0.4}$. (c) $T$ against $\alpha$ with the type-I chimera as the initial state, $D=10^{-5}$. (d) $T$ against $D$ with the type-II chimera as the initial state, $\alpha=1.47$ and the data is fitted to be $T\sim D^{-0.56}$ (the blue line).}
\end{figure}

\begin{figure}
  \centering
  \includegraphics[width=3.5in]{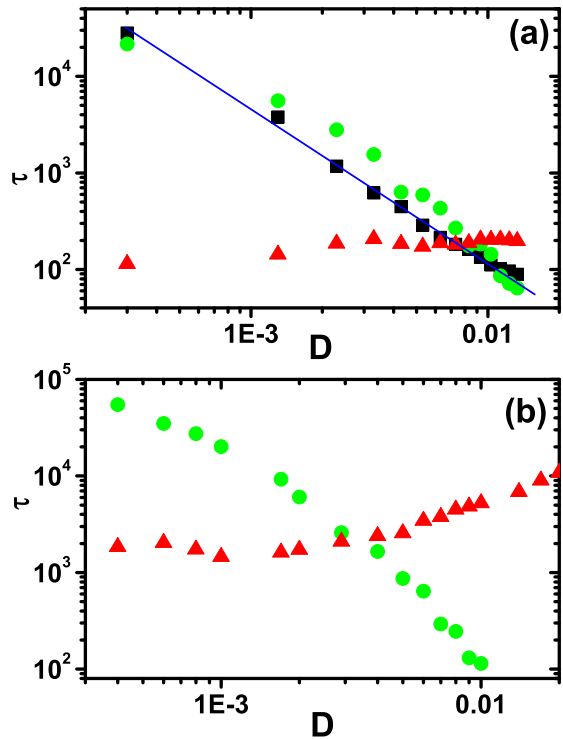}
  \caption{\label{fig5}The mean lifetimes of the synchronous state (black), the type-I chimera (red), and the type-II chimera (green) are plotted against $D$. $\alpha=1.38$ in (a) and $\alpha=1.47$ in (b). }
\end{figure}
The response of dynamical states to the mobility of agents may be related to the basin stability \cite{men13}. The basin stability is a method to account for the stability of a dynamical state by using its attraction basin but not the linear stability analysis. Generally, the basin stability of a dynamical state is strong if the fraction of initial conditions leading to the state is high. There are two competing processes when a dynamical state is subjected to perpetual perturbation, the perturbation drives the system away from the state and even to exit its attraction basin while the nonlinear stability draws the system within the attraction basin back towards it. In this view, how a system owning multistable states reacts to perpetual noise is determined by the basin stabilities of these dynamical states and the strength of perpetual noise. The stronger the basin stability of a state, the more possibly the state to be an absorbing state. The stronger the perpetual noise, the more possibly the system to be taken away from a dynamical state. For the model Eqs.~(\ref{eq1}-\ref{eq3}), the basin stabilities of the type-I and the type-II chimeras increase with $\alpha$ while the one of the synchronous state decreases with $\alpha$. Considering that the global parameter $R$ for the type-I chimera is in between the synchronous state and the type-II chimera, it seems that it is easier for the type-II chimera to transit to the type-I chimera than to the synchronous state, which is also observed in extensive simulations. We first consider the case with weak mobility. For $\alpha=1.38$, the basin stability of the synchronous state is much stronger than that of the type-I chimera and, the type-II chimera is not sustainable, which leads to the synchronous state to be the only absorbing one. For $\alpha=1.47$, the type-II chimera is stable and the results in Fig.~\ref{fig3}(d) suggest that the type-I and the type-II chimeras have comparable basin stabilities. In comparison with the synchronous state, the type-I and the type-II chimeras have weaker basin stabilities, which results in that initial type-I chimera is absorbed by the synchronous state. However, the type-II chimera is sustained since the type-I chimera plays a role of buffer to keep the type-II chimera away from the synchronous state. For the case with intermediate mobility, the noise is strong enough to take the system out of the attraction basin of any dynamical state. As a result, we find the frequent jumps of the system among different states for $\alpha=1.38$. It is also interesting to note that, the system is just only between the two chimeras for $\alpha=1.48$ and stays around the type-I chimera for $\alpha=1.54$, which suggest that the driving away from dynamical states by perpetual perturbation and drawing back to the state by its nonlinear stability contributes to the response of the system in a collective way but not in a simple way. When noise is sufficiently strong, the perpetual noise dominates and it drives the system away from the three states. Under such strong noise, the effect of nonlinear stability can hardly be seen.

To get a clearer view on the response of the system to the mobility of agents, we investigate the transient time $T$ of a given initial state before it is captured by the synchronous state for weak mobility (small $D$). Bearing the Brownian motions of agents in mind, we know that transient time $T$ may not be the same in different simulation runs even for the same initial conditions. For this sake, we record the transient time for those realizations leading the system to the absorbing synchronous state. The transient time $T$ for each $D$ is averaged over $20$ realizations. Figures~\ref{fig4}(a) and (b) show $T$ against $D$ for $\alpha=1.38$ and $\alpha=1.47$ with the type-I chimera as initial conditions, respectively. Interestingly, $T$ seems to scale with $D$ roughly as $T\sim D^{-\gamma}$. Fitting the curve with a linear function in the double-log plot, we find that the exponent $\gamma\simeq0.4$ seems to be insensitive to $\alpha$. To be mentioned, the scaling relation between T and $D$ is fitting well just for weak mobility. For $D>4\times10^{-4}$ at $\alpha=1.38$ and for $D>4\times10^{-6}$ at $\alpha=1.47$, the synchronous state is competing with the jumping between different dynamical states, which results in the violation of the scaling relation. Actually, the transient time $T$ here amounts to the first passage time of the coupled oscillators to the boundary of the attraction basin of the type-I chimera. For a Brownian particle in a homogeneous space, the first passage time of the particle to leave a given domain scales with $D$ with an exponent $-1$. $\gamma=0.4$ in Figs.~\ref{fig4}(a) and (b) suggests that both perpetual perturbation due to the mobility of agents and the nonlinear stability of the type-I chimera collectively contribute to the scaling behavior of $T$ against $D$. The transient time $T$ against $D$ tends to deviate from the scaling law at large $D$. The comparison between Fig.~\ref{fig4}(a) and (b) also shows that the transient time $T$ at large $\alpha$ is much larger than that at small $\alpha$, behind which is that the basin stability of the synchronous state (the chimera states) decreases (increase) with $\alpha$. Then, we investigate the transient time $T$ against $\alpha$ at $D=10^{-5}$. As shown in Fig.~\ref{fig4}(c), $T$ increases with $\alpha$. When $\alpha$ goes down below $1.33$, $T$ decreases to $0$ since chimera states become unstable at sufficient low $\alpha$. On the other hand, $T$ shows a sharp rise and large fluctuation at around $\alpha=1.49$ beyond which the synchronous state is no longer the absorbing state. As shown in Fig.~\ref{fig4}(d) where the type-II chimera is chosen to be initial states, the transient time $T$ to the synchronous state also exhibits a scaling law with $D$. The exponent $\gamma\simeq0.56$ is different from that with the type-I chimera as initial conditions, which further supports the collective contributions to the transient time from the mobility of agents and the nonlinear stability of the current dynamical state. Comparing with the case with the type-I chimera as initial conditions, the case with the type-II chimera as initial conditions shows longer transient time and stronger fluctuation, both of them result from the facts that the route from the type-II chimera to the synchronous state has to involve the type-I chimera.

For intermediate mobility, the coupled phase oscillators jump among different dynamical states and the lengths of the time sections holding a certain dynamical state may fluctuate greatly. We are interested in the jumping dynamics which may be characterized by the mean lifetimes $\tau$ of different dynamical states. To calculate $\tau$, we set two $R$, $R_1=0.9$ and $R_2=0.4$. The dynamical state is thought to be in the synchronous state when $R>R_1$, in the type-II chimera when $R<R_2$, and the type-I chimera when $R_2<R<R_1$. To reduce the fake jumps among these states, we monitor the window averaged $R$. The coupled phase oscillators jump among the synchronous state and the two chimera states at $\alpha=1.38$, and Fig.~\ref{fig5}(a) presents $\tau$ against $D$ for these states. With $D$ increasing, we find that the mean lifetimes of both the synchronous state and the type-II chimera decrease. On the contrary, the mean lifetime of the type-I chimera increases with the mobility $D$ and saturates at large $D$ such as $D=10^{-2}$. The observations that $\tau$ tends to zero for both the synchronous state and the type-II chimera and that $\tau$ stays at around hundreds of time units for the type-I chimera at large $D$ indicate that the jumps among the three states become frequent with $D$ increase. Then we consider the jump dynamics at $\alpha=1.47$ where the jumps occur mainly between the two chimera states. We treat the time sections holding the type-I chimera as those with $R>R_2$ and holding the type-II chimera otherwise. The mean lifetimes against $D$ are presented in Fig.~\ref{fig5}(b). Increasing the mobility always leads the rise (fall) of the lifetime of the type-I (the type-II) chimera. The abnormal behavior of the lifetime of the type-I chimera that $\tau$ is much high for $D<10^{-3}$ hints that the existence the synchronous state in some time sections cannot be ignored. Additionally, Fig.~\ref{fig5} shows that the time sections supporting either the synchronous state or the type-II chimera do not exist for high mobility.

\section{Conclusion}
In this work, we considered $N$ agents performing Brownian motions on a ring. Each agent carries a phase oscillator and interacts with others with a kernel function depending on the distance between them. When agents are motionless, the model reduces to a standard one in the field of chimera state and allows for three dynamical states, the synchronous state, the chimera state with one coherent domain (the type-I), and the chimera state with two coherent domains (the type-II). The motion of agents impacts on the model dynamics by the perpetual disturbances induced by the change of relative positions among agents. We found that the response of the coupled phase oscillators to the movement of agents depends on both the phase lag $\alpha$ and the agent mobility $D$. For low mobility, the type-I chimera state transits to the synchronous state for $\alpha$ more less than $\pi/2$ and becomes an absorbing state otherwise. For intermediate mobility, the coupled oscillators randomly jump among different dynamical states depending on $\alpha$. We also investigated the statistical properties in these different dynamical regimes. We found the scaling laws between the transient time and the mobility for low mobility and presented the relations between the mean lifetimes of different dynamical states and the mobility for intermediate mobility. In this work, we did not discuss the situation with sufficiently high mobility. Uriu et. al. \cite{uriu13} have shown that mobile coupled oscillators behave as a mean field system at high mobility of oscillators and Petrungaro et. al. \cite{Peru10} have showed that chimera states disappear for very high mobility and the global synchronization dominates the system. For the model with Eqs.~(\ref{eq1}-\ref{eq3}) in this work, the synchronous state with all oscillators being in phase is no longer the solution to the model [see the inset in fig2(a)]. Though the chimera dynamics is lost at sufficiently high mobility, the seemingly disorder states with $R\neq1$ appear.

\begin{acknowledgements}
This work was supported by National Natural Science Foundation of China (Nos. 11575036, 11805021 and 11505016).
\end{acknowledgements}

Compliance with ethical standards. The authors declare that they have no conflict of interest concerning the publication of this manuscript.


\begin{thebibliography}{}

\bibitem{kura} Kuramoto, Y., Battogtokh, D.: Coexistence of coherence and incoherence in nonlocally coupled phase oscillators. Nonlinear Phenom. Complex Syst. \textbf{5}, 380-385 (2002)
\bibitem{abra} Abrams, D.M., Strogatz, S.H.: Chimera states for coupled oscillators. Phys. Rev. Lett. \textbf{93}, 174102 (2004)
\bibitem{lai09} Laing, C.R.: The dynamics of chimera states in heterogeneous Kuramoto networks. Physica D \textbf{238}, 1569 (2009)
\bibitem{mot10} Motter, A.E.: Nonlinear dynamics: Spontaneous synchrony breaking. Nat. Phys. \textbf{6}, 164-165 (2010)
\bibitem{zhu12} Zhu, Y., Li, Y., Zhang, M., Yang, J.: The oscillating two-cluster chimera state in non-locally coupled phase oscillators. EPL \textbf{97}, 10009 (2012)
\bibitem{pan15} Wu, Z., Cheng, H., Feng, Y., Li, H., Dai, Q., Yang, J.: Chimera states in bipartite networks of FitzHugh-Nagumo oscillators,  Front. Phys. \textbf{13}, 130503 (2018)
\bibitem{mart13} Martens, E.A., Thutupalli, S., Fourri\`{e}re,  A., Hallatschek, O.: Chimera states in mechanical oscillator networks. Proc. Nat. Acad. Sci. USA \textbf{110}, 10563-10567 (2013)
\bibitem{tins12} Tinsley, M.R., Nkomo, S., Showalter, K.: Chimera and phase-cluster states in populations of coupled chemical oscillators. Nat. Phys. \textbf{8}, 662-665 (2012)
\bibitem{hage12} Hagerstrom, A.M., Murphy, T.E., Roy, R., H\"{o}vel, P., Omelchenko, I., Sch\"{o}ll E.: Experimental observation of chimeras in coupled-map lattices. Nat. Phys. \textbf{8}, 658-661 (2012)
\bibitem{cheng18} Cheng, H., Dai, Q., Wu, N., Li, H., Yang, J.: Chimera states in nonlocally coupled phase oscillators with biharmonic interaction. Commun. Nonlinear Sci. Numer. Simul \textbf{56}, 1 (2018)
\bibitem{gav18} Li, X., Bi, R., Sun, Y., Zhang, S., Song, Q.: chimera states in Gaussian coupled map lattices, Front. Phys. \textbf{13}, 130502 (2017)
\bibitem{lai18} Xu, H., Wang, G., Huang, L, and Lai, Y: Chaos in Dirac Electron Optics: Emergence of a Relativistic Quantum Chimera. Phys. Rev. Lett. \textbf{120}, 124101 (2018)
\bibitem{omel11} Omelchenko, I., Maistrenko, Y., H\"{o}vel, P., Sch\"{o}ll, E.,: Loss of coherence in dynamical networks: Spatial chaos and chimera states. Phys. Rev. Lett. \textbf{106}, 234102 (2011)
\bibitem{omel15} Omelchenko, I., Zakharova, A., H\"{o}vel, P., Siebert, J., Sch\"{o}ll, E.: Nonlinearity of local dynamics promotes multi-chimeras. Chaos \textbf{25}, 083104 (2015)
\bibitem{omel13} Omelchenko, I., Omelchenko, O.E., H\"{o}vel, P., Sch\"{o}ll, E.: When nonlocal coupling between oscillators becomes stronger: Patched synchrony or multichimera states. Phys. Rev. Lett. \textbf{110}, 224101 (2013)
\bibitem{hiz14} Hizanidis, J., Kanas, V., Bezerianos, A., Bountis, T.: Chimera states in networks of nonlocally coupled Hindmarsh-Rose neuron models. Int. J. Bifurcat. Chaos \textbf{24}, 1450030 (2014)
\bibitem{sak06} Sakaguchi, H.: Instability of synchronized motion in nonlocally coupled neural oscillators. Phys. Rev. E \textbf{73}, 031907 (2006)
\bibitem{oml15} Olmi, S., Martens, E.A., Thutupalli, S., Torcini, A.: Intermittent chaotic chimeras for coupled rotators. Phys. Rev. E \textbf{92}, 030901(R) (2015)
\bibitem{yel14} Yeldesbay, A., Pikovsky, A., Rosenblum, M.: Chimeralike states in an ensemble of globally coupled oscillators. Phys. Rev. Lett. \textbf{112}, 144103 (2014)
\bibitem{set14} Sethia, G.C., Sen, A.: Chimera states: The existence criteria revisited. Phys. Rev. Lett. \textbf{112} (2014)
\bibitem{cha14} Chandrasekar, V.K., Gopal, R., Venkatesan, A. Lakshmanan, M.: Mechanism for intensity-induced chimera states in globally coupled oscillators. Phys. Rev. E \textbf{90}, 062913 (2014)
\bibitem{pre15} Premalatha, K., Chandrasekar, V.K., Senthilvelan, M., Lakshmanan, M.: Impact of symmetry breaking in networks of globally coupled oscillators. Phys. Rev. E \textbf{91}, 052915 (2015)
\bibitem{lai15} Laing, C.R.: Chimeras in networks with purely local coupling. Phys. Rev. E \textbf{92}, 050904(R) (2015)
\bibitem{ber16} Bera, B.K., Ghosh, D., Lakshmanan, M.: Chimera states in bursting neurons. Phys. Rev. E \textbf{93}, 012205 (2016)
\bibitem{mart10a} Martens, E.A, Laing, C.R., Strogatz, S.H.: Solvable model of spiral wave chimeras. Phys. Rev. Lett. \textbf{104}, 044101 (2010)
\bibitem{gu13} Gu, C., St-Yves, G., Davidsen, J.: Spiral wave chimeras in complex oscillatory and chaotic systems. Phys. Rev. Lett. \textbf{111}, 134101 (2013)
\bibitem{pana13} Panaggio, M.J., Abrams, D.M.: Chimera states on a flat torus. Phys. Rev. Lett. \textbf{110}, 094102 (2013)
\bibitem{pana15} Panaggio, M.J., Abrams, D.M.: Chimera states on the surface of a sphere. Phys. Rev. E \textbf{91}, 022909 (2015)
\bibitem{zhuy14} Zhu, Y., Zheng, Z., Yang, J.: Chimera states on complex networks. Phys. Rev. E \textbf{89}, 022914 (2014)
\bibitem{jiang16} Jiang, X., Abrams, D.M.: Symmetry-broken states on networks of coupled oscillators. Phys. Rev. E\textbf{93} 052202 (2016)
\bibitem{sem16} Semenova, N., Zakharova, A., Anishchenko, V.,  Sch\"{o}ll, E.: Coherence-Resonance Chimeras in a Network of Excitable Elements. Phys. Rev. Lett. 117, 014102 (2016)
\bibitem{dai18} Dai, Q., Zhang, M., Cheng, H., Li, H., Xie, F., Yang, J.: From collective oscillation to chimera state in a nonlocally coupled excitable system. Nonlinear dyn. \text{91}, 1723 (2018)
\bibitem{mot17} Cho, Y.S., Nishikawa, T., Motter, A.E.: Stable Chimeras and Independently Synchronizable Clusters. Phys. Rev. Lett. \textbf{119}, 084101 (2017)
\bibitem{dai18b} Dai, Q., Yang, K., Cheng, H, Li, H., Xie, F., Yang, J.: Chimera states in nonlocally coupled bicomponent phase oscillators: From synchronous to asynchronous chimeras. arXiv:1808.03220v1 [nlin.AO]
\bibitem{sac00} Sachtjen, M.L., Carreras, B.A., Lynch, V.E.: Disturbances in a power transmission system. Phys. Rev. E \textbf{61}, 4877 (2000)
\bibitem{olf07} Olfati-Saber, R., Fax, J.A., Murray, R.M.: Consensus and cooperation in networked multi-agent systems. Proc. IEEE \textbf{95}, 215 (2007)
\bibitem{onn07} Onnela, J.P., Saramaki, J., Hyvonen, J., Szabo, G., Lazer, D., Kaski, K., Kertesz, J., Barabasi, A.L.: Structure and tie strengths in mobile communication networks. Proc. Nat. Acad. Sci. USA \textbf{104}, 7332 (2007)
\bibitem{maj17} Majhi, S., Ghosh, D.: Amplitude death and resurgence of oscillation in networks of mobile oscillators. EPL \textbf{118}, 40002 (2017)
\bibitem{shen13} Shen, C., Chen, H., Hou, Z.: Mobility-enhanced signal response in metapopulation networks of coupled oscillators. EPL \textbf{102}, 38004 (2013)
\bibitem{fra08} Frasca, M., Buscarino, A., Rizzo, A., Fortuna, L., Boccaletti S.: Synchronization of Moving Chaotic Agents. Phys. Rev. Lett. \textbf{100}, 044102 (2018)
\bibitem{gom13} G\'{o}mez-Garde\~{n}es, J., Nicosia, V., Sinatra, R., Latora, V.: Motion-induced synchronization in metapopulations of mobile agents. Phys. Rev. E \textbf{87}, 032814 (2013)
\bibitem{eom16} Eom, Y., Boccaletti, S., Caldarelli, G.: Concurrent enhancement of percolation and synchronization in adaptive networks. Sci. Rep. \textbf{6}, 27111 (2016)
\bibitem{fuj11} Fujiwara, N., Kurths, J., D\'{i}az-Guilera, A.: Synchronization in networks of mobile oscillators. Phys. Rev. E \textbf{83}, 025101(R) (2011)
\bibitem{pri13} Prignano, L., Sagarra, O., D\'{i}az-Guilera, A.: Tuning Synchronization of Integrate-and-Fire Oscillators through Mobility. Phys. Rev. Lett. \textbf{110}, 114101 (2013)
\bibitem{bea17} Beardo, A., Prignano, L., Sagarr,a O., D\'{i}az-Guilera, A.: Influence of topology in the mobility enhancement of pulse-coupled oscillator synchronization. Phys. Rev. E \textbf{96}, 062306 (2017)
\bibitem{Petr17} Petrungaro, G., Uriu, K., Morelli, L.G.: Mobility-induced persistent chimera states, Phys. Rev. E \textbf{96}, 062210 (2017)
\bibitem{wol11} Wolfrum, M., Omelchenko, O.E.: Chimera states are chaotic transients. Phys. Rev. E \textbf{84}, 015201(R) (2011)
\bibitem{men13} Menck, P.J., Heitzig, J., Marwan, N., Kurths, J.: How basin stability complements the linear-stability paradigm. Nat. Phys. \textbf{9}, 89 (2013)
\bibitem{uriu13} Uriu, K., Ares, S., Oates, A.C., Morelli, L.G.: Dynamics of mobile coupled phase oscillators. Phys. Rev. E \textbf{87}, 032911 (2013)
\bibitem{Peru10} Peruani, F., Nicola, E.M., Morelli, L.G.: Mobility induces global synchronization of oscillators in periodic extended systems. New J. Phys. \textbf{12}, 093029 (2010)



\end{thebibliography}
\end{document}